\documentclass[11pt,twoside]{article}
\usepackage[american]{babel}
\usepackage{times,subeqnarray}
\usepackage{url}
\newif\myifpdf
\ifx\pdfoutput\undefined
   \usepackage[dvips]{graphicx}
\else
   \pdfoutput=1        
    \usepackage[pdftex]{graphicx}
\fi
\usepackage{apatitlepages}
\usepackage{setspace} 
\usepackage{psydraft}
\usepackage[natbibapa]{apacite}   

\usepackage{enumitem}
\usepackage{subcaption}
\usepackage{csquotes}

\myifpdf
  \DeclareGraphicsExtensions{.pdf,.eps,.png,.jpg,.mps,.tif}
\fi

\def\myheading{ Predictive Learning in Speech }

\def\mytitle{Statistical Learning in Speech: A Biologically Based Predictive Learning Model}

\def\myauthor{John Rohrlich and Randall C. O'Reilly\\
  Department of Psychology, Computer Science, and Center for Neuroscience \\
  University of California Davis \\
  1544 Newton Ct\\
  Davis, CA 95618\\
  {\small jarohrlich@ucdavis.edu}}

\def\mynote{
Thanks to Tom Hazy for comments on a draft version.

R. C. O'Reilly is Chief Scientist at eCortex, Inc., which may derive indirect benefit from the work presented here.

Supported by: ONR grants N00014-20-1-2578, N00014-19-1-2684/ N00014-18-1-2116, N00014-18-C-2067, N00014-17-1-2961, N00014-15-1-0033.

All data and materials will be available at \url{https://github.com/ccnlab/statlearn} upon publication.}

\def\myabstract{
Infants, adults, non-human primates and non-primates all learn patterns implicitly, and they do so across modalities. The biological evidence supports the hypothesis that the mechanism for this learning is general but computationally local. We hypothesize that the mechanism itself is predictive error-driven learning. We build on recent work that proposed a biologically plausible model of error backpropagation learning which proposes that higher order thalamic nuclei provide a locale for a temporal difference between top-down predictions and an actual event outcome. Our neural network based on that work also models the auditory cortex hierarchy of core, belt and parabelt and the caudal-rostral axis within regions. We simulated two studies showing statistical learning in infants, a seminal study using synthesized speech and a more recent study using human speech. Before simulating these studies the network was trained on spoken sentences from the TIMIT corpus to emulate infants experience listening to random speech. The implemented neural network, learning only by predicting the next brief speech segment, learned in both simulations to predict in-word syllables better than next-word syllables showing that prediction could be the basis for word segmentation and thus statistical learning.
}

\begin{document}
\bibliographystyle{apacite}

\titlesepage{\mytitle}{\myauthor}{\mynote}{\myabstract}

\pagestyle{myheadings}

A significant question for cognitive science, especially regarding early human development, is how we learn from raw sensory experiences. How is it that infants learn the patterns and regularities in their environment? It is widely thought that some aspects of this learning are well-captured by the phenomenon of \emph{statistical learning}, which has been documented across visual \citep{FiserAslin01, FiserAslin02, KirkhamSlemmerJohnson02} and auditory senses \citep{SaffranAslinNewport96, HauserNewportAslin01}, in non-human primates \citep{ConwayChristiansen01, HauserNewportAslin01} and across species \citep{ToroTrobalon05}, suggesting a fundamental mechanism.  An important feature of statistical learning is that it happens automatically and without awareness of the product of the learning \citep{FrenschRunger03}, consistent with a putative role in driving the foundational learning across sensory domains and a critical role in the development of language, speech, motor planning, music, and so on.  One of the most widely-studied forms of statistical learning involves infants learning to segment continuous speech streams based on the co-occurrence statistics of phonemes \citep{SaffranAslinNewport96,  GrafEstesLew-Williams15}, which is the focus of this paper.

At a computational level, \emph{predictive learning} provides an appealing mechanism driving this phenomenon of statistical learning. Studies going back to \citet{Elman90} have shown that a simple recurrent network (SRN) learns structure by predicting.  Studies by \citet{CleeremansMcClelland91} and \citet{Elman91} showed that SRNs, i.e. models that use recent information to predict what is next, can learn an artificial grammar. \citet{CairnsShillcockChaterEtAl97} found that an SRN could learn, to an imperfect degree, to segment words from a large corpus of transcribed English speech. \citet{ChristiansenAllenSeidenberg98} employing an SRN, and adding additional cues for better prediction, also modeled speech segmentation, adding lexical stress and utterance boundaries (two to six words) to the phoneme information. Two additonal studies modeling segmentation with SRNs \citep{GirouxRey09, PlautVandeVelde17} looked at the role of chunking and part-words when learning to segment speech. All of these studies showing that prediction is a good candidate for how infants learn to segment speech and learn the rules of grammar.

We build on this work to explore a recent proposal for the biological basis of a closely-related form of predictive learning, based on unique properties of the corticothalamic circuitry interconnecting the deep layers of the neocortex and the higher-order thalamic nuclei such as the pulvinar \citep{OReillyRussinZolfagharEtAl21}.  In this model, predictive learning starts immediately within the sensory processing pathways of the brain, e.g., predictive learning of the visual world takes place within the lower-level visual-specific pathways, as contrasted with a possible alternative model where this learning is only operating at a higher, more abstract level.  In this paper, we investigate whether this model can account for relevant empirical data in the domain of statistical learning in infants in the context of the widely-studied speech segmentation paradigm  \citep{SaffranAslinNewport96, GrafEstesLew-Williams15}.

In our simulations we found \textit{in-word} syllable transitions were better predicted than \textit{next-word} syllable transitions, with a model that learns solely from prediction error on the raw auditory speech stream.  Given the biological and modality specificity of our model, it is also important that our model accurately reflect the known auditory cortical and thalamic pathways and patterns of connectivity. As such, the neural network we implemented captures both known auditory neuroscience and corticothalamic computation supporting predictive learning.  Thus, across these different levels of analysis, our simulations support the proposal that such a sensory-specific, biologically-based predictive learning mechanism could be the basis for development of word segmentation in speech, and, by extension, likely other forms of statistical learning. 

\subsection{Statistical Learning}

The term statistical learning was first used by \citet{SaffranAslinNewport96} and the first experiments on infant statistical learning were focused on word segmentation \citep{SaffranAslinNewport96, AslinSaffranNewport98}. Those experiments provide evidence that infants can segment fluent speech on the basis of the co-occurrence probability between adjacent syllables, a statistical feature called a transitional probability. There are of course other cues in speech that help one detect word boundaries but these were absent in these experiments. Since those first experiments, studies using the same paradigm have demonstrated statistical learning with tones \citep{ConwayChristiansen06}, in infants and adults \citep{SaffranJohnsonAslinEtAl99, GrafEstesLew-Williams15}, and in newborns \citep{BulfJohnsonValenza11}. Statistical learning has also been shown with visual stimuli \citep{FiserAslin02, BulfJohnsonValenza11} and in serial reaction time tasks \citep{HuntAslin01}. What is the biology supporting this type of learning?

Though there are not many studies addressing the biology underlying statistical learning they are consistent in showing enhanced activation in modality specific areas when listening to or viewing repeating, as opposed to random, streams. Studies of repeating speech streams \citep{CunilleraCamaraToroEtAl09, KaruzaNewportAslinEtAl13, ElmerValizadehCunilleraEtAl21} show this enhanced activation in the superior temporal gyrus and the superior portion of the ventral premotor cortex when compared to non-patterned streams. When processing repeating visual sequences enhanced activation occurs in high-level visual cortex \citep{Turk-BrowneSchollChunEtAl09}. These results suggest that local computations in separate cortical areas are responsible for statistical learning.

This hypothesis of local computation is further supported by research showing individual differences in statistical learning across modalities and across stimuli within modality. If the biology supporting statistical learning was not in low level modality specific processing we would not expect to see such differences. \citet{ConwayChristiansen06} found that participants in an artificial grammar learning paradigm can learn two different sets of statistical regularities simultaneously when stimuli are from different modalities, and even within one modality when the two streams differ in a major perceptual dimension, again supporting the hypothesis that statistical learning is governed by local computation. \citet{FrostArmstrongSiegelmanEtAl15} hold this view and hypothesize that statistical learning is a set of domain-general computational principles that work within each modality and within the constraints of that modality and each brain region. We believe that the most important of these mechanisms is predictive learning, based on detailed properties of the neocortex and thalamus, as reviewed next.

\subsection{Auditory Neuroscience}

The biological mechanism proposed by \citet{OReillyRussinZolfagharEtAl21} is a specific form of predictive error-driven learning based on distinctive patterns of connectivity between the neocortex and the higher-order nuclei of the thalamus. For audition these higher-order nuclei are the anterodorsal and posterodorsal areas of the medial geniculate body (MGB) \citep{ShermanGuillery06, UsreySherman18}. The hypothesis is that learning is driven by the difference between top-down predictions, generated by numerous weak projections into the thalamic relay cells (TRCs) in the higher-order nuclei, and the actual outcomes supplied by sparse, strong driver inputs from lower areas. The driver inputs originate in cortical layer 5 and are from bursting neurons (5IBs) so they are brief and periodic, firing roughly every 100 ms, i.e. alpha frequency. Thus, the prediction error is a temporal difference in activation states over the higher-order thalamic nuclei, from an earlier prediction to a subsequent burst of outcome. This temporal difference can drive local synaptic changes throughout the neocortex, supporting a biologically-plausible form of error backpropagation that improves the predictions over time \citep{OReilly96,AckleyHintonSejnowski85,HintonMcClelland88,BengioMesnardFischerEtAl17,WhittingtonBogacz19,LillicrapSantoroMarrisEtAl20}.

The organization of the human auditory cortex, in part based on what is known of non-human primate auditory cortex, is generally held to have three hierarchically organized regions known as the \emph{core, belt} and \emph{parabelt} \citep{KaasHackett00}. Each of these regions is further divided, with the core regions having at least two areas, the belt region perhaps seven and the parabelt region two. Connection patterns suggest a forward flow of information from core to belt to parabelt and also a flow caudal to rostral \citep{HackettdelaMotheCamalierEtAl14, ScottLecceseSaleemEtAl17}. Feedback connections also exist along both axes.

Thalamic input to auditory cortex comes primarily from the MGB. Input to the core areas, A1 and R, coming from the ventral division of the MGB while belt and parabelt regions, both caudal and rostral, receive predominantly from the dorsal divisions of the MGB \citep{HackettStepniewskaKaas98, MotheBlumellKajikawaEtAl06a, LeeSherman10}. This difference in thalamic source is central to the predictive learning mechanism, described in more detail in the next section.

The core areas A1 and R are both tonotopically organized but there are functional differences. One difference that is well supported is the longer latency responses in R compared to A1 \citep{CamalierDAngeloSterbing-DAngeloEtAl12, BendorWang08}. Another difference in these core areas is in the tracking of amplitude modulations. Area A1 can track fast acoustic amplitude modulations, on the order of 20 -- 30 ms, whereas R can only track amplitude modulations of 100 ms or more \citep{ScottMaloneSemple11}. Thus, the caudal core auditory cortex responds quickly to sound onsets and tracks fast amplitude modulations accurately, whereas the rostral auditory cortex responds more slowly and tracks longer amplitude modulations (see \citet{JasminLimaScott19} for more on rostral-caudal contributions to auditory processing). These differences are consistent with the rostral portion being responsible for phoneme analysis \citep{JasminLimaScott19, DeWittRauschecker12, CamalierDAngeloSterbing-DAngeloEtAl12}.

Along the other axis, the hierarchy of core, belt and parabelt, it has been found that while core fields respond best to tones and noise, belt and parabelt respond best to more complex stimuli. Recording studies in macaque have shown that there is a broadening of neural response to wider frequency ranges moving from core to non-core areas \citep{RauscheckerTianHauser95, RecanzoneGuardPhan00}. This is consistent with imaging studies in macaques and humans showing that non-core areas respond best to speech \citep{BinderFrostHammekeEtAl00, LiebenthalBinderSpitzerEtAl05b}.

\section{Computational Model}

\begin{figure}[h]
  \centering\includegraphics[width=6in]{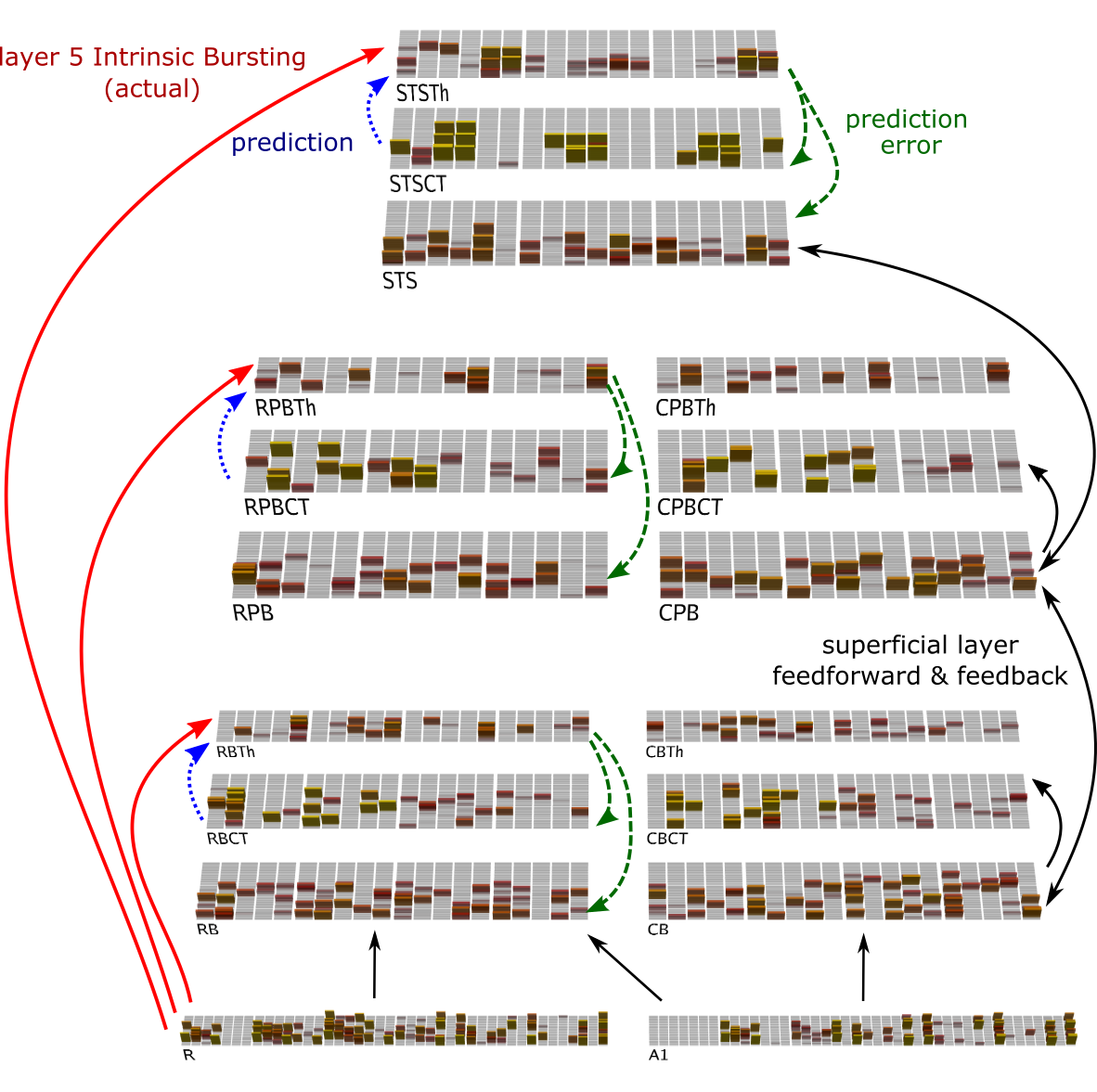}
  \caption{\footnotesize Predictive learning model showing the \textbf{prediction} onto the higher order thalamic nuclei from cortical layer 6 (e.g. RPBCT $\rightarrow$ RPBTh; dotted blue arrows), \textbf{actual} auditory signal from cortical layer 5 intrinsic bursting neurons (e.g. R $\rightarrow$ RPBTh; solid red arrows), and the \textbf{prediction error} (e.g. RPBTh $\rightarrow$ RPBCT and RPBTh $\rightarrow$ RPB; dashed green arrows). Caudal predictive pathways (not shown) are the same as rostral. A few of the typical feedforward and feedback pathways (solid black arrows) are shown for caudal layers (not shown for rostral layers). Caudal to rostral feedforward and feedback pathways also not shown. Appendix includes a full list of projections. Layers: caudal belt (CB), rostral belt (RB), caudal parabelt (CPB), rostral parabelt (RPB), and superior temporal sulcus (STS)}
  \label{fig.netview}
\end{figure}

Figure \ref{fig.netview} shows the implemented neural network used for the simulations, both the areas and the most important pathways. The model includes the hierarchy of core, belt and parabelt regions, both caudal and rostral, and the superior temporal sulcus (STS) receiving from both the caudal parabelt (CPB) and rostral parabelt (RPB). Our simplified cortical model has two belt areas rather than the likely seven \citep{KaasHackett00}. Moving from core to belt to parabelt in the model, neuron tuning is broader in both the frequency and the time dimensions, with projections largely being constrained within the caudal and rostral areas. However, there are feedforward and feedback projections between caudal and rostral superficial layers with the feedforward being stronger than the feedback consistent with the anatomy described in the previous section. Auditory input is to the core areas A1 and R and is maintained throughout each 100 ms trial. There is no feedback to the core areas in this model.  All of this is consistent with a traditional model of the auditory system.

The model goes further by including the higher order thalamic areas (e.g. RBTh, RPBTh) and an implementation of hypothesized predictive circuity that drives implicit learning. To make the working of the predictive mechanism more accessible, the belt, parabelt and superior temporal sulcus cortical areas are implemented with separate neurons for the superficial layers, (1--4) and the deeper layers (5--6), which we refer to as corticothalamic (CT suffix). The thalamic areas are grouped with corresponding cortical layers rather than shown more anatomically.

\begin{figure}[h]
  \centering\includegraphics[width=6in]{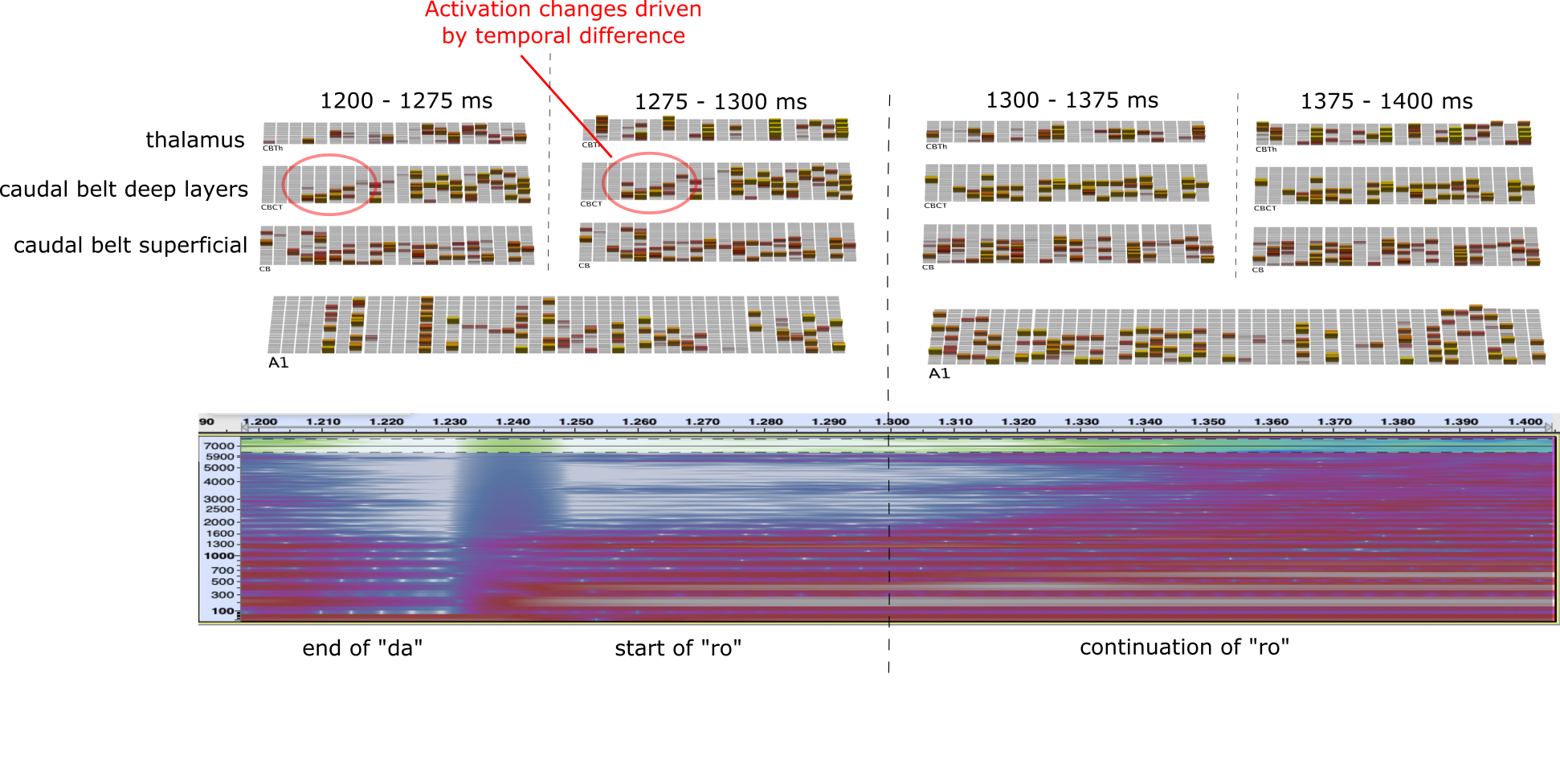}
    \caption{\footnotesize A 200 ms sample of corticothalamic flow (2 alpha cycles) showing the prediction activation on the thalamus and the activation from the layer 5IB cells (showing A1 layers, other areas are similar). This temporal difference is the error which drives learning. During the first 75 ms of any 100 ms period (e.g. 1200 ms to 1275 ms) a prediction is projected from deep layers of cortical areas onto higher order thalamic nuclei. In the later part of the 100 ms period (1275 ms to 1300 ms) cortical layer 5 bursting from the actual auditory stimulus drives the thalamus. The temporal difference (i.e. the prediction error) is projected back onto superficial and deep cortical layers, which can be seen in subtle activation changes when comparing with the activations at the 75 ms point.}
  \label{fig.corticothalamic}
\end{figure}

The flow of information between superficial layers, deep layers and thalamic areas is shown in Figure \ref{fig.netview}. Core areas receive the auditory signal which is passed via superficial layer pathways to other auditory cortical areas. Each area's deep layers, receiving in part from superficial, maintain the context and via layer 6 project a prediction onto higher order thalamic nuclei. In the model, prediction takes place during the first 75 ms of an alpha cycle (minus phase). During the last 25 ms of the alpha cycle layer 5IB cell bursting occurs (plus phase). This burst from the core areas is the current true input. The difference in higher order thalamic activation between the prediction and the actual is the prediction error that is projected back to the cortical areas which drives learning. Figure \ref{fig.corticothalamic} shows a snapshot of actual model activation for a 200 ms period early in training. The \enquote{delay} allows time for the prediction to develop and thus the temporal difference which implicitly encodes a prediction error. This is a critical point, if the belt and parabelt were receiving direct thalamic input from the ventral portion of the MGB they would have the ground truth immediately and there would be no temporal difference from prediction to outcome.

Speech sequences for pretraining, training and testing were input to the network 150 ms per trial with a stride of 100 ms. The 100 ms (i.e. alpha cycle) stride is based on the biology described previously. The 150 ms input on each trial provides an overlap in processing that prevents artifacts from abrupt transitions. To encourage invariance, a random duration of silence, 0--25 ms, was added at runtime at the start of each sequence. This was done for both training and testing (i.e. holdout) sequences.

The 150 ms of sound was processed in time windows of 25 ms, with a 10 ms overlap, through a Fourier Transform to calculate the power spectrum. This was processed through a Mel filter bank of 42 filters covering the range of 20 Hz to 6000 Hz. Finally, the Mel filter-bank output was convolved with 7 gabor filters which is the input to the A1 and R (i.e. core) layers of the network.

As mentioned in the Auditory Neuroscience section there are processing and response differences between caudal and rostral core areas and in the many belt regions as well. We did not attempt to model these differences in detail as the focus of this work is on the biological mechanism supporting statistical learning but we recognize that differences in neuron response timing, amplitude tracking, frequency selectivity and so on will undoubtedly interact with predictive learning. We did implement one difference, however, different size gabors for the caudal and rostral input to A1 and R respectively. The gabor filters used to process the auditory signal to R were narrower than those for A1 to capture the more fine grained transients involved in phoneme discrimination consistent with the rostral portion of the auditory cortex being part of the ventral stream of auditory processing \citep{DeWittRauschecker12,JasminLimaScott19}.

\section{Simulation 1}

Simulation 1 was modeled on the experiments reported in \citet{SaffranAslinNewport96} and \citet{AslinSaffranNewport98}. In those experiments infants listened to an artificial language of four nonsense words where each word in the corpus was composed of three consonant-vowel pairs (CVs). The synthesized CVs were generated at a constant rate with no acoustic cues to word boundaries. The only clue to word boundaries was frequency of co-occurrence of CVs.  In each experiment a familiarization phase, a few minutes of listening to the stream of CVs, was followed by a test phase using isolated words, two from the four word corpus and two that were composed of the last CV of one word followed by the first two CVs of another corpus word. Infants responses to the \textit{whole-words} vs \textit{part-words} showed reliable discrimination based on transitional probability. Our simulation used the same corpus and showed that the model predicts the neural activations in the next 100 ms time period in line with transitional probabilities.

\section{Method}

\subsection{Stimuli}

The synthesized sound files from the experiment by  \citet{SaffranAslinNewport96} were not available. A small sample of the original speech generation was available along with comments from one author on how the sound sequence was generated (J.R. Saffran, personal communication, August 13, 2020). This information guided the reproduction of the sounds using GnuSpeech \citep{HillTaube-SchockManzara17}. The corpus was composed of the four trisyllable words, \textit{daropi, golatu, pabiku and tibudo}, a total of 12 syllables. To create the speech files randomized lists of the 12 CVs were passed to GnuSpeech. Each resulting WAV file was edited to create 12 separate files, one for each CV. The trisyllable words were created by concatenating from random instances of each CV. Finally, 24 sound files, all permutations of four words, were generated by concatenating the trisyllable words. The duration of each sequence was approximately 3730 ms, with a mean of 3.25 CVs per second. Syllable duration ranged from 167 ms (\textit{ti}) to 447 ms (\textit{ro}). See Appendix for details on the creation of the stimuli.

\subsection{Train and Test}

To simulate infant experience, hearing a wide variety of sentences and voices, we pretrained on a subset of the TIMIT corpus \citep{GarofoloLamelFisherEtAl93}. TIMIT is a corpus of phonemically and lexically transcribed speech of American English speakers of different sexes and dialects. We used the 450 \enquote{SX} sentences spoken by the 128 female speakers. Training on the TIMIT corpus was done for 50 epochs with each epoch containing 10 randomly chosen sentences. After every 10 epochs of pretraining we tested the model with the artificial corpus we would later train on, to have a learning baseline.

Each of the 25 pretraining runs on the TIMIT corpus was followed by training on the four trisyllable corpus. Training on the trisyllable corpus was for 10 epochs, nine sequences per epoch. The nine sequences (36 trisyllable words) were chosen randomly each epoch from 20 of the 24 total sequences. Four random sequences were held back each run for testing. Testing was done after every epoch for this training. A new random seed was used for each of the 25 runs.

\subsection{Measuring Performance}

If the model is learning to predict it should be able to predict in-word syllables, which are fully predictable, better than next-word syllables, where there are three possibilities. We measured prediction during the first 100 ms of every CV other than the first of each sequence. Specifically, we compared the thalamic layer activations at the 75 ms point (i.e. the end of the prediction phase) with the activations of the thalamic layer at the 100 ms point (i.e. 25 ms after the 5IB neurons begin the periodic burst). The cosine difference between these values is the measure of prediction.

\begin{figure}[h]
  \centering\includegraphics[width=6in]{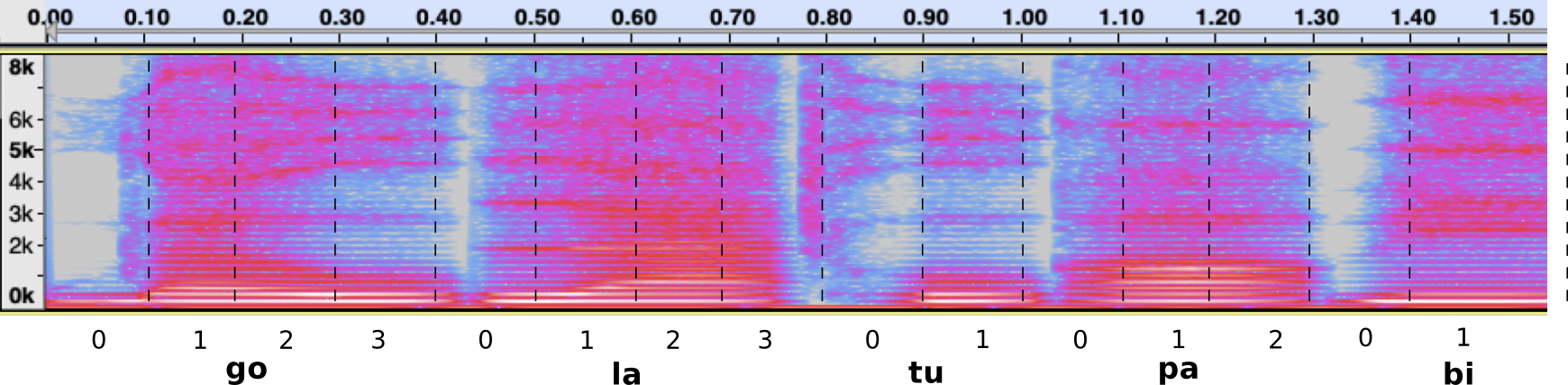}
  \caption{\footnotesize A spectrogram showing the first 1500 ms of the 4 word sequence \enquote{golatupabikutibudodaropi}. Prediction was measured on the first 100 ms of each CV except the first of the sequence, in this case \enquote{go}. The second CV, \enquote{la} starts at about 430 ms so we use the 100 ms period from 400--500 ms. If there had been less than 70 ms of \enquote{la} in the 400--500 ms period the next 100 ms would have been considered the first subsegment of \enquote{la}. That situation occurs with the third CV, \enquote{tu}, which starts at about 770 ms. We consider the 800--900 ms region the first 100 ms of \enquote{tu}.}
  \label{fig.spectro}
\end{figure}

Though the measurement was done every 100 ms, coinciding with the shift in input, CVs varied in duration, as they do in all human speech, and additionally were offset by random silence at the start of each sequence. This means that the 100 ms period may include the tail of one CV along with the start of another or that a portion of the start of a CV is missed. This is true for both in-word and next-word measurements so we looked at the in-word / next-word comparison rather than any absolute value. Figure \ref{fig.spectro} is a 1500 ms sample from a spectrogram showing the 100 ms segments for \enquote{golatupabi} with a 0 ms duration of silence at the sequence start. The first 400 ms comprise the CV  \enquote{go} though a bit of the CV sound falls beyond 400 ms. The first segment of \enquote{la} includes the tail of \enquote{go}. We used a 70\% threshold for determining the CV. Adding silence or sampling from a difference sequence may change the number of 100 ms segments for a CV. Additionally the same CV will vary in length for different speakers and also in the synthesized sounds though to a smaller degree.

\section{Results and Discussion}

\begin{figure}[h]
  \centering\includegraphics[width=6in]{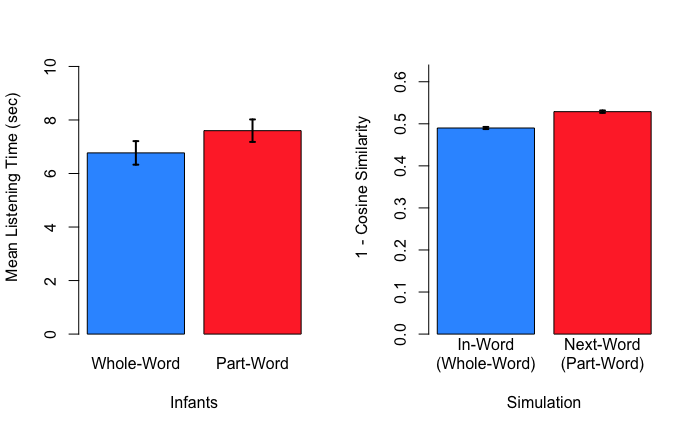}
  \caption{\footnotesize Results from \cite{SaffranAslinNewport96} experiment alongside those of Simulation 1, both having statistically significant differences between conditions. Note that the measure in the experiment was listening time while for the simulation it was prediction accuracy. Both charts show the result after 180 words. Error bars indicate SEs.}
  \label{fig.saffranExpVsSim}
\end{figure}

Data was evaluated using the cosine similarity value for each 100 ms period at the start of each syllable. A linear mixed model, which handles both fixed and random effects, was used. The first analysis looked at the test results after the 5th epoch of training, the point at which the model had \enquote{heard} the same number of words as the infants in the \citet{SaffranAslinNewport96} experiment. The model fixed effects were, the factor of main interest, \textit{condition}: in-word transition and next-word transition; \textit{layer}: the five sets of neurons representing specific auditory cortical areas. \textit{Run}, equivalent to one subject, was treated as a random variable. Comparing the full model to one without the condition factor showed a significant difference: \(\chi ^2(1) = 181.44, p < 0.001\), prediction being better for in-word transitions. Figure \ref{fig.saffranExpVsSim} shows the result from \citet{SaffranAslinNewport96} experiment and the simulation at a comparable point. Note that for the \citet{SaffranAslinNewport96} experiment the important point is that if infants listened to either group of words significantly longer it would mean that they had extracted the crucial sequence information. It was the part-words they listened to longer. For the simulation the direction of the difference is important and the difference was in the expected direction, in-word transitions were predicted better than next-word transitions.

\begin{figure}[h]
  \centering\includegraphics[width=6in]{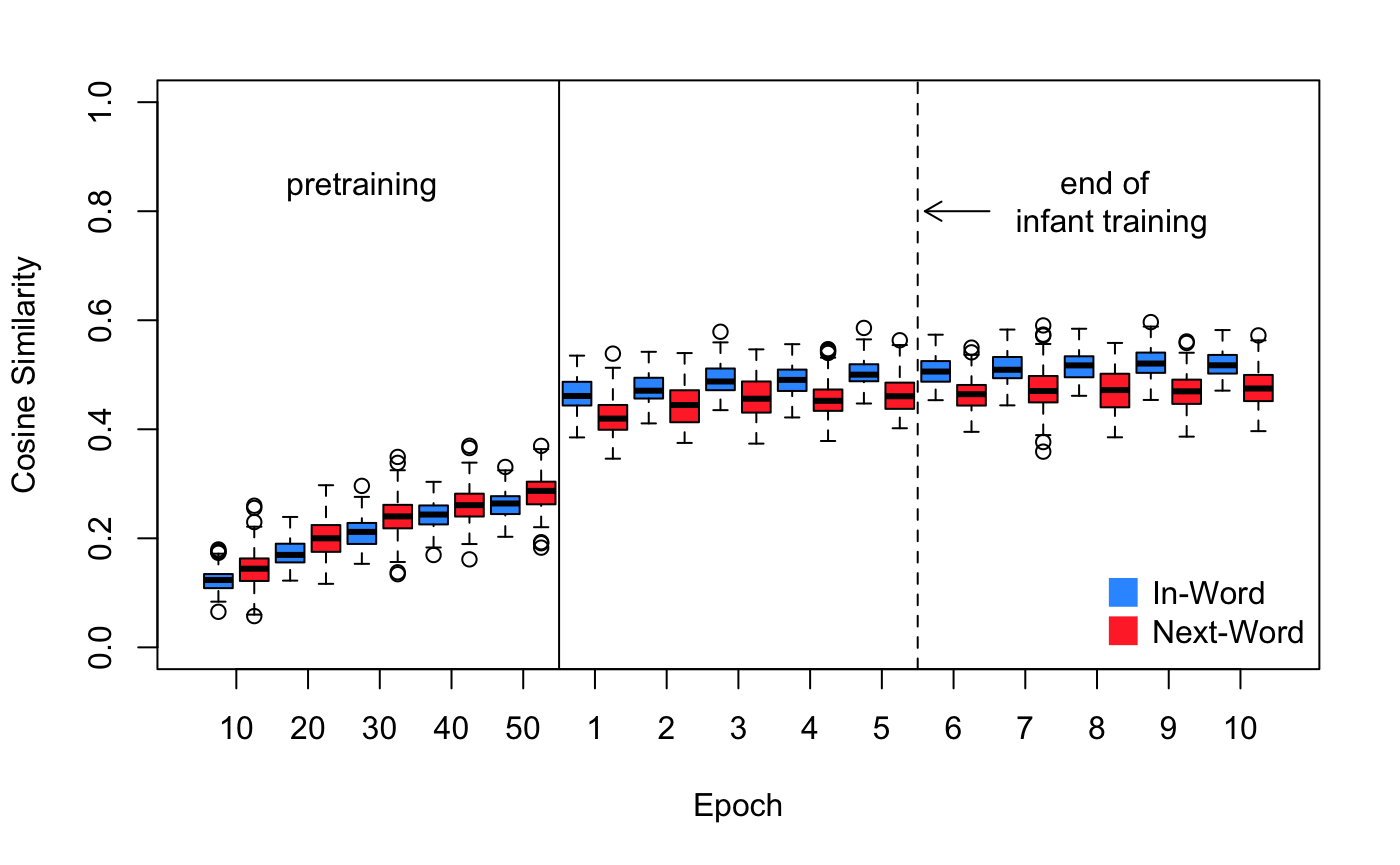}
\caption{\footnotesize A box plot showing the model predicts in-word transitions better than next-word transitions. Testing was done every 10 epochs during pretraining and after each epoch during training. The measure is the similarity of activations at the end of the \enquote{minus} phase of each trial (prediction) and the actual neuron activations at the end of the \enquote{plus} phase. The infants in the \cite{SaffranAslinNewport96} experiment listened to 180 words. The model was trained on 36 words each epoch.}
  \label{fig.saffranPrePostTesting}
\end{figure}

\begin{figure}[h!]
  \centering\includegraphics[width=6in]{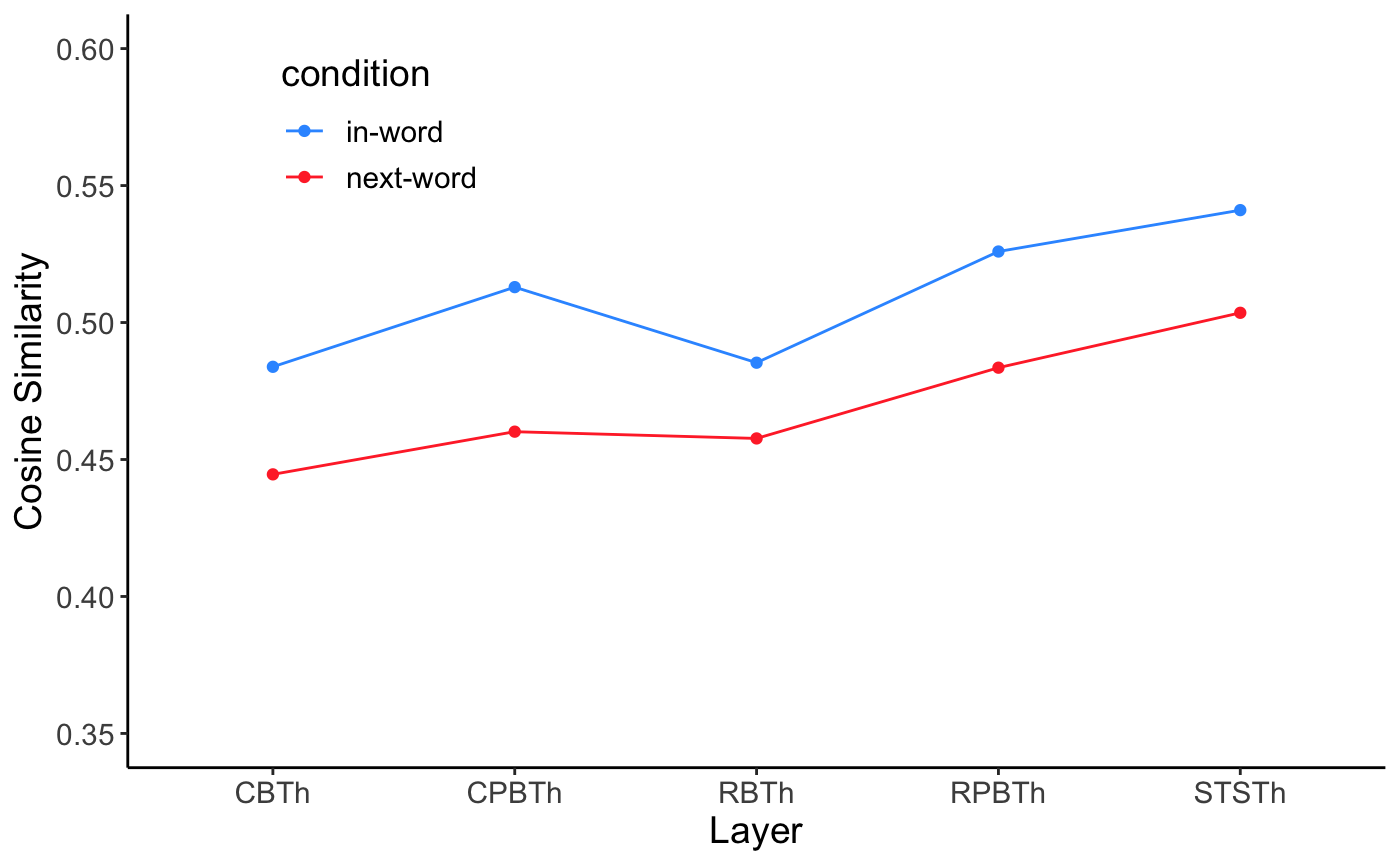}
  \caption{\footnotesize Mean cosine similarity values by condition and by layer showing higher layers predicting better than lower and rostral layers predicting better than caudal. There was an interaction of layer and condition so the planned comparisons were analyzed separately by condition, see Table \ref{tab.sim1comparisons} for the statistical analysis. The cosine similarity measures are on the thalamic (Th) area corresponding to each layer. Layers: caudal belt (CB), rostral belt (RB), caudal parabelt (CPB), rostral parabelt (RPB), and superior temporal sulcus (STS)}
  \label{fig.saffranInteraction}
\end{figure}

Results for the full simulation, rather than epoch 5 in isolation, are shown in Figure \ref{fig.saffranPrePostTesting}. For this analysis epoch was added as a fixed factor, pretraining and training were analyzed separately. Pretraining on the TIMIT corpus showed significant learning across epochs, \(\chi ^2(1) = 1913.7 , p < 0.001\), and a significant effect of condition, \(\chi ^2(1) = 219.59 , p < 0.001\), with next-word transitions being predicted better than in-word transitions. Analysis of training on the artificial language showed a significant condition effect across epochs, \(\chi ^2(1) = 1397.8 , p < 0.001\), as was the case for epoch 5 analyzed by itself. In-word transitions were predicted better than next-word transitions, as expected. The analysis also showed a significant effect for layer \(\chi ^2(4) = 1547.3 , p < 0.001\) and the layer by condition interaction \(\chi ^2(4) = 78.63 , p < 0.001\) (see Figure  \ref{fig.saffranInteraction}).

For comparison, we also ran the model without the TIMIT pretraining. A comparable level of prediction performance was achieved after 25 epochs, with in-word prediction better than next-word prediction, as it was with pretraining (see Figure \ref{fig.saffranNoPretrainTesting} in Appendix).

\begin{table}[h]
\begin{subtable}[]{0.45\textwidth}
\centering
\begin{tabular}{|l|l|l|l|}
\hline
Comparison & df & t-value & p-value \\
\hline
STS - RPB & 2485 & 13.84 & p \textless{} .001 \\
STS - CPB & 2485 & 7.44 & p \textless{} .001 \\
RPB - RB  & 2485 & 14.32  & p \textless{} .001 \\
CPB - CB  & 2485 & 19.99 & p \textless{} .001 \\
RPB - CPB & 2485 & 6.40 & p \textless{} .001 \\
RB - CB   & 2485 & 0.74  & p = 0.46 \\
\hline
\end{tabular}
\caption{In-Word Condition}
\end{subtable}
\begin{subtable}[]{0.45\textwidth}
\centering
\begin{tabular}{|l|l|l|l|}
\hline
Comparison & df & t-value & p-value \\
\hline
STS - RPB & 2485 & 21.37 & p \textless{} .001 \\
STS - CPB & 2485 & 9.86 & p \textless{} .001 \\
RPB - RB  & 2485 & 7.68  & p \textless{} .001 \\
CPB - CB  & 2485 & 12.72 & p \textless{} .001 \\
RPB - CPB & 2485 & 11.50 & p \textless{} .001 \\
RB - CB   & 2485 & 6.46  &  p \textless{} .001 \\
\hline
\end{tabular}
\caption{Next-Word Condition}
\end{subtable}
\caption{\footnotesize Table showing the six planned comparisons for Simulation 1 of layers predicting activations (shown separately for in-word and next-word conditions because of the interaction). The measure of prediction is the cosine difference between predicted and actual activations on the thalamus. The prediction on the thalamus comes from the deep layers of the cortex areas, the layers with the CT suffix in Figure \ref{fig.netview}. All layer to layer comparisons were in the expected direction for the in-word condition, higher layers predicting better than lower and rostral predicting better than caudal. The results were the same for the next-word condition with the exception of the RB to CB comparison which was nonsignificant. Layers: superior temporal sulcus (STS), rostral parabelt (RPB), caudal parabelt (CPB), rostral belt (RB) and caudal belt (CB)}
\label{tab.sim1comparisons}
\end{table}

By training the model on one corpus, TIMT, and testing on another, our four nonsense word corpus, we see that the model is learning to predict the activations from auditory input generally. That next-word transitions were better predicted than in-word is surprising, we expected no difference, but a four word corpus is a very small test set. Testing after training on the four word corpus showed the expected opposite result, better prediction of in-word (i.e. fully predictable) transitions. All of the six planned comparisons were in the expected direction for the in-word condition, five of the six for the next-word condition (see Table \ref{tab.sim1comparisons}). The STS layer predicted activations better than parabelt layers, parabelt layers predicted better than belt layers and rostral layers predicted better than caudal. This is in line with the two axis organization of the auditory cortex and will be discussed further in the General Discussion.

\section{Simulation 2}

Simulation 2 was modeled on experiment 1 reported in \citep{GrafEstesLew-Williams15}. This experiment is different from the \citet{SaffranAslinNewport96} experiment, Simulation 1, primarily in the use of human rather than synthesized speech and the use of multiple voices, rather than a single voice. Further, it is a more difficult task because the next-word transition probability was 50\% rather than 33\%. In the \citet{GrafEstesLew-Williams15} experiment infants, 8- and 10-month-olds, listened to a language of four nonsense words where each word in the corpus was composed of two syllables. The six minutes of speech heard by the infants was edited from the recordings of eight female speakers. Each speaker recorded monotone three-syllable sequences that incorporated all of the possible coarticulation contexts present in the language and the two syllable words were created by excising and splicing the middle syllable of the three-syllable sequences. This technique reduced the chance of speakers introducing supplemental word boundary cues. To create variation, syllables from each speaker were spliced together in sequences that ranged from 10 to 20 syllables, a duration of 3 s to 7 s, and the order of speakers was randomized.

\section{Method}
\subsection{Stimuli}

We were able to obtain an original sound file used in the experiments of \citet{GrafEstesLew-Williams15}. It was for version \enquote{A} of the artificial language (\textit{timay, dobu, gapi, and moku}). The counterbalancing language was not available. The 3 minutes of speech, heard twice in the original experiment, were subdivided into 18 segments of approximately 10 seconds each.

\subsection{Train and Test}

As in simulation 1, speech sequences were input to the network 150 ms per trial with a stride of 100 ms. To add variability to the stimuli a short random duration of silence, 0 - 25 ms, was added at runtime at the start of each sequence. Sequences were processed in the same manner as for Simulation 1. Each epoch was composed of 6 sequences, about 60 seconds of speech, and the simulation ran for 10 epochs. The model was run 25 times, holding out three sound files for testing on each run. Each simulation began with a new random seed. Performance was measured as in Simulation 1.

\section{Results and Discussion}

\begin{figure}[h]
  \centering\includegraphics[width=6in]{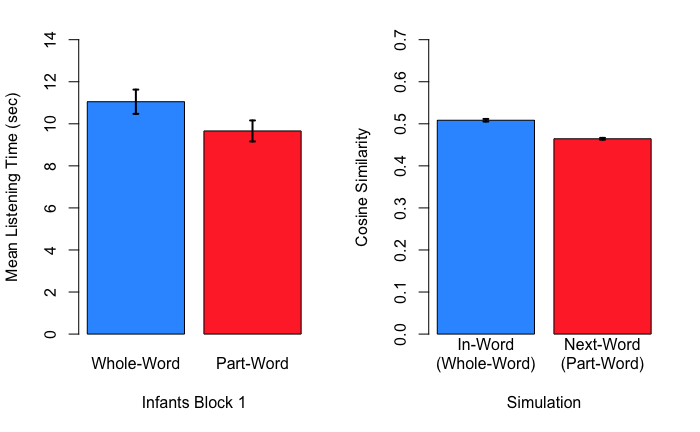}
  \caption{\footnotesize Results from \cite{GrafEstesLew-Williams15} experiment alongside those of Simulation 2, both having statistically significant differences between conditions. Note that the measure in the experiment was listening time while for the simulation it was prediction accuracy. Both charts show the result after 6 minutes of \enquote{listening} words. Error bars indicate SEs.}
  \label{fig.grafestesExpVsSim}
\end{figure}

Performance of the model was measured in the same manner as for Simulation 1. The cosine similarity values for each 100 ms period at the start of a syllable were analyzed with a linear mixed model. We look first at the results of testing after the 6th epoch of training, the point at which the model had heard six minutes of speech, the same duration of speech as the infants in the \citet{GrafEstesLew-Williams15} experiment. The model fixed effects were layer: the five sets of neurons representing specific auditory cortical areas; condition: in-word transition or next-word transition. Run, equivalent to one subject, was treated as a random variable. Comparing the full model to one without the condition factor showed a significant difference: \(\chi ^2(1) = 172.36, p < 0.001\). Figure \ref{fig.grafestesExpVsSim} shows the result from \citet{GrafEstesLew-Williams15} experiment and the simulation at a comparable point. In the \citet{GrafEstesLew-Williams15} experiment infants listened significantly longer to the familiar words (i.e. whole-words). In the simulation in-word transitions, equivalent to the whole-word condition, were predicted better than next-word transitions.

\begin{figure}[h]
  \centering\includegraphics[width=6in]{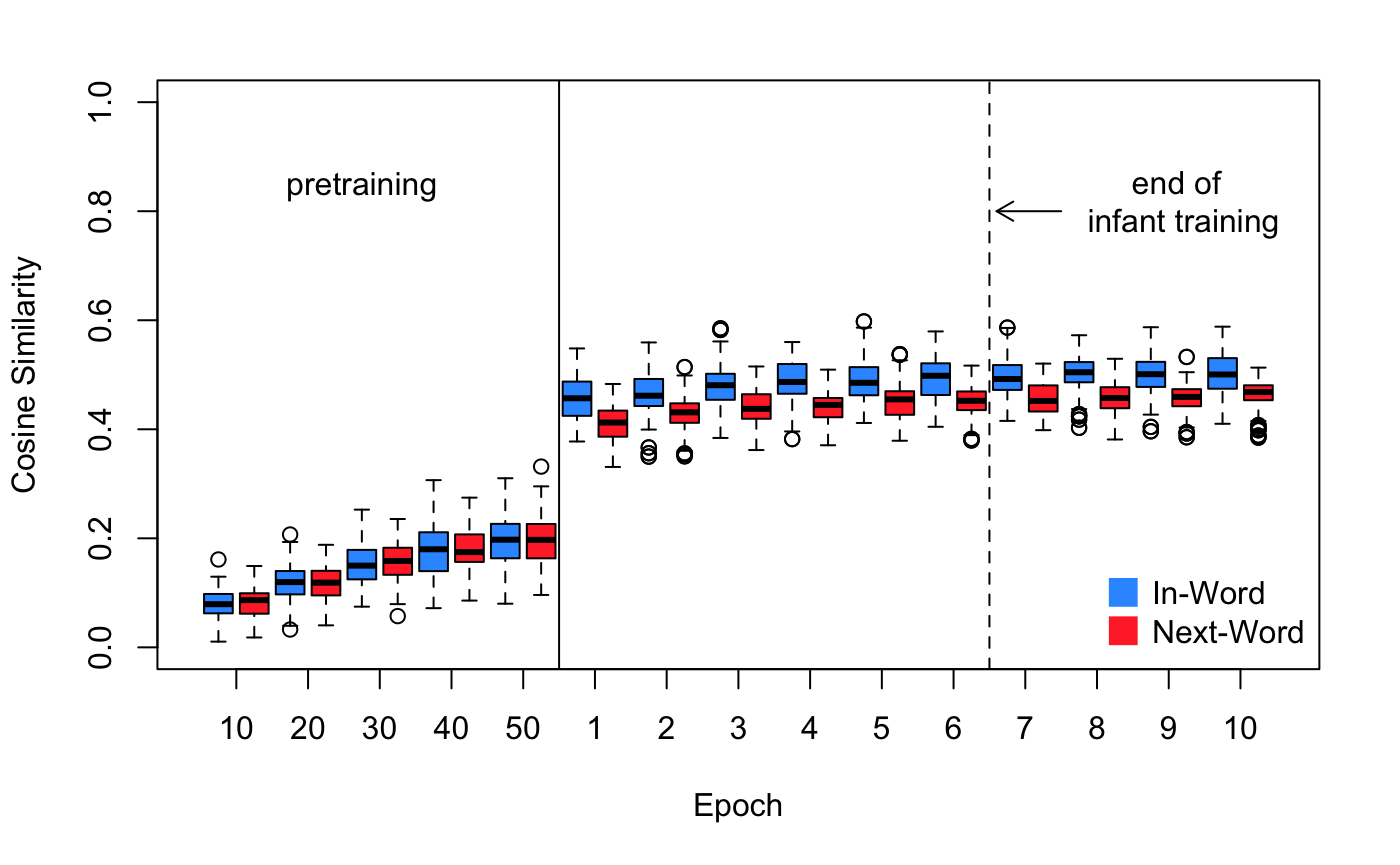}
  \caption{\footnotesize A box plot showing the model predicts in-word transitions better than next-word transitions. Testing was done every 10 epochs during pretraining and after each epoch during training. The measure is the similarity of activations at the end of the \enquote{minus} phase of each trial (prediction) and the actual neuron activations at the end of the \enquote{plus} phase. The infants in the \citet{GrafEstesLew-Williams15} experiment listened to 6 minutes of speech. The model was trained with one minute of speech each epoch.}
  \label{fig.grafestesPrePostTesting}
\end{figure}

\begin{figure}[h!]
  \centering\includegraphics[width=6in]{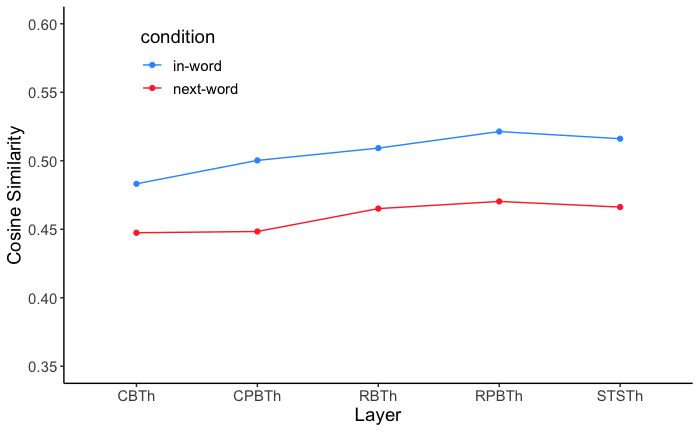}
  \caption{\footnotesize Mean cosine similarity values by condition and by layer showing parabelt layers predicting better than belt layers and rostral layers predicting better than caudal. STS prediction was not better than RPB prediction as expected. There was an interaction of layer and condition so the planned comparisons were analyzed separately by condition, see Table \ref{tab.sim2comparisons} for the statistical analysis. The cosine similarity measures are on the thalamic (Th) area corresponding to each layer. Layers: caudal belt (CB), rostral belt (RB), caudal parabelt (CPB), rostral parabelt (RPB), and superior temporal sulcus (STS)}  \label{fig.grafestesInteraction}
\end{figure}

Results for the full simulation, rather than epoch 6 in isolation, are shown in Figure \ref{fig.grafestesPrePostTesting}. For this analysis epoch was added as a fixed factor, pretraining and training were analyzed separately. Pretraining on the TIMIT corpus showed significant learning across epochs, \(\chi ^2(1) = 1570.6 , p < 0.001\), and also a small effect of condition, \(\chi ^2(1) = 10.27 , p < .01\), in-word prediction better than next-word prediction. Analysis of training on the artificial language showed a significant condition effect across epochs, \(\chi ^2(1) = 1604.9 , p < 0.001\), as it did looking at epoch 6 separately. In-word transitions were predicted better than next-word transitions, as expected and as in Simulation 1. The analysis also showed a significant effect for layer \(\chi ^2(4) = 480.60 , p < 0.001\) and for the layer by condition interaction \(\chi ^2(4) = 38.28 , p < 0.001\) (see Figure \ref{fig.grafestesInteraction}). The planned comparisons, the same as those for Simulation 1, were in the expected direction (see Table \ref{tab.sim2comparisons}) with a couple of exceptions, STS did not predict better than RPB though it did predict better than CPB. Also, the CPB / CB comparison showed no difference. These results are in line with the two axis organization of the auditory cortex as they were in Simulation 1.  For comparison, we also ran the model without the TIMIT pretraining. A comparable level of prediction performance was achieved after 20 epochs, with in-word prediction better than next-word prediction, as it was with pretraining (see Figure \ref{fig.grafestesNoPretrainTesting} in Appendix). 

\begin{table}[h!]
\begin{subtable}[]{0.45\textwidth}
\centering
\begin{tabular}{|l|l|l|l|}
\hline
Comparison & df & t-value & p-value \\
\hline
STS - RPB & 2485 & 7.28 & p \textless{} .001 \\
STS - CPB & 2485 & -2.40 & p = 0.016 \\
RPB - RB  & 2485 & 7.84  & p \textless{} .001 \\
CPB - CB  & 2485 & 5.55 & p \textless{} .001 \\
RPB - CPB & 2485 & 9.68 & p \textless{} .001 \\
RB - CB   & 2485 & 11.96 & p \textless{} .001 \\
\hline
\end{tabular}
\caption{In-Word Condition}
\end{subtable}
\begin{subtable}[]{0.45\textwidth}
\centering
\begin{tabular}{|l|l|l|l|}
\hline
Comparison & df & t-value & p-value \\
\hline
STS - RPB & 2485 & 8.19 & p \textless{} .001 \\
STS - CPB & 2485 & -1.89 & p = 0.059 \\
RPB - RB  & 2485 & 0.43  & p = 0.67 \\
CPB - CB  & 2485 & 2.40 & p = 0.016 \\
RPB - CPB & 2485 & 10.07 & p \textless{} .001 \\
RB - CB   & 2485 & 8.10  &  p \textless{} .001 \\
\hline
\end{tabular}
\caption{Next-Word Condition}
\end{subtable}
\caption{\footnotesize Table showing the six planned comparisons for Simulation 2 of layers predicting activations (shown separately for in-word and next-word conditions because of the interaction). The measure of prediction is the cosine difference between predicted and actual activations on the thalamus. The prediction on the thalamus comes from the deep layers of the cortex areas, the layers with the CT suffix in Figure \ref{fig.netview}. As expected RPB prediction was better than RB, CPB prediction better than CB. Both rostral areas predicted better than caudal. STS did predict better than CPB but not RPB. Layers: superior temporal sulcus (STS), rostral parabelt (RPB), caudal parabelt (CPB), rostral belt (RB) and caudal belt (CB)}
\label{tab.sim2comparisons}
\end{table}

\section{General Discussion}

The fundamental role of prediction in learning is receiving increasing attention and support. The simulations reported in this paper show that prediction is a possible basis for infant word segmentation and statistical learning more generally. The two simulations replicating experiments in infant learning both showed that a neural network model based on a biologically plausible form of error-driven learning and prediction, grounded in the anatomy of the auditory thalamocortical pathways, learns to predict next syllable transitions from speech. This learning followed prelearning on a widely used corpus of varied content and voice that was unrelated to the artificial language that the model was later trained and tested on. This prelearning served to simulate infants exposure to a wide variety of voices and sentences. In both simulations the model predicted in-word syllable transitions better than less predictable next-word transitions. The infant experiments reported in \citet{SaffranAslinNewport96} and \citet{GrafEstesLew-Williams15} also showed a clear difference between in-word and next-word conditions but not a reliable direction of effect. In the \citet{SaffranAslinNewport96} experiment infants listened longer to the \enquote{unfamiliar} words (i.e. the next-word condition) while in the \citet{GrafEstesLew-Williams15} experiment infants listened longer to the familiar words in Block 1 and the unfamiliar in Block 2 of testing. Although our model has no direct equivalent of listening time, we hypothesize that prediction success and prediction errors contribute to attentional engagement.

More generally, these results support the hypothesis that predictive learning, using the hypothesized biologically-based mechanism, underlies statistical learning in infants. The nearly immediate statistical learning effects after pretraining suggest that it has established generalizable phoneme-like representations that provide an efficient basis for subsequent learning about phoneme-level co-occurrence statistics. These results add to our earlier  work in the visual modality \citep{OReillyRussinZolfagharEtAl21} showing the development of abstract categorization purely from predictive learning. The hypothesis of predictive error-driven learning in the thalamocortical loop was first advanced in \citet{OReillyWyatteRohrlich14}.

A couple of aspects of the specific results reported deserve discussion. In both simulations in-word transitions were better predicted than next-word transitions, but the prediction did not reach 100 percent for in-word, though the transition was fully predictable. We believe the ceiling on prediction is largely due to measurement on 100 ms boundaries. The varying length of CVs plus the random silence means that the 100 ms period where the cosine similarity is measured will rarely coincide with the actual first 100 ms of the CV auditory signal. We also need to consider the model itself. The syllables spoken in Simulation 1 had durations in the 167 ms to 447 ms range while in Simulation 2 the range was 241 ms to 422 ms with the larger portion of the syllable duration being the vowel sound. This can lead to ambiguity in predicting. Consider just two words from Simulation 1, \enquote{pabiku} and \enquote{daropi}. Using only the last 100 ms or so of the first CV would make \enquote{bi} as good a guess as \enquote{ro}. Part of what needs to be learned in learning to predict is what information is useful and for what period of time. Improving this aspect of learning is planned for future work.

The interaction of prediction accuracy, layer and condition also deserves some discussion. In the auditory cortex incoming information generally flows from core to belt to parabelt and from caudal regions to more rostral, with regions further along the hierarchy receiving a broader signal in both time and frequency. Thus individual neurons in the parabelt, for example, have a longer time horizon to use in predicting than say a belt area neuron. Likewise, the rostral areas have more information on which to base predictions than caudal areas. The results from both simulations followed these expectations with parabelt layers predicting better than belt and rostral layers predicting better than caudal. STS receives from both caudal and rostral and predicted better than both in Simulation 1 and better than caudal but not rostral in Simulation 2. How this aspect of the model matches the biology is something that might be testable in the future.

Although we attempted to capture major features of auditory cortex, including both caudal and rostral areas, there are further details that were not fully captured in our model. There is evidence, for example, that the core rostral region, R, exhibits longer response latencies than the core caudal region, A1 \citep{ScottMaloneSemple11, CamalierDAngeloSterbing-DAngeloEtAl12, NourskiSteinschneiderMcMurrayEtAl14}. It has also been reported that synchronization of spike discharges to modulations of stimulus amplitude and frequency are integrated over quite different time windows in A1 (20--30 ms) and R (100 ms) \citep{ScottMaloneSemple11}. To broadly reflect these differences in our model, we processed the signal for core areas A1 and R differently, using narrower filtering in R to capture shorter sound transients that would be useful in phoneme differentiation.

There are multiple cues involved in learning to segment words, and evidence suggests that the favored cues change during development, which may provide further clues into the nature of the learning process. The experiments with infants that we simulated specifically controlled the speech, both synthesized and human, to eliminate word boundary cues other than transition probability, cues such as prosody, relating to linguistic stress, and phonotactic rules, governing language specific sequences of sounds. Infants use these cues as well as transition probability. \citet{JohnsonJusczyk01} pitted prosodic cues against statistical cues and found that 8-month-olds weighed speech cues more heavily than statistical cues. \citet{ThiessenSaffran03} found that 7-month-olds attended more to statistical cues than to stress cues but 9-month-olds ignored statistical cues and segmented based on syllable stress. These development differences could be modeled and further test our predictive learning hypothesis.

In conclusion, the simulations described in this paper extend the work on prediction error-driven learning in thalamocortical circuits into the auditory domain and show that the same mechanism that allows the visual system to learn to categorize shapes \citep{OReillyRussinZolfagharEtAl21} can do statistical learning in the auditory circuits. That important learning is done via prediction has been put forth by \citet{Elman90, Friston05, GeorgeHawkins09, RaoBallard99} to name a few.  Our demonstration that predictive learning on raw auditory signals can result in sensitivity to statistical regularities in the speech stream that mirrors those of infants provides an initial foundation for future work on predictive learning of more sophisticated and abstract features of the speech stream.  How far can purely predictive learning on raw sensory input go toward providing a capable foundation for higher-level speech comprehension, and motor production?  These are questions we are eager to address in future research.

\section{Appendix}

All of the materials described here, including the speech synthesis, speech files and the computational model are all available on our github account at: \url{https://github.com/ccnlab/statlearn}. The {\em emergent} simulation environment is at: \url{https://github.com/emer/leabra} which contains extensive documentation and examples that can be run in Python or the Go language.  The best place to start in understanding computationally how the predictive learning model works is with the FSA model described in \citet*{OReillyRussinZolfagharEtAl21}, which is available at: \url{https://github.com/emer/leabra/tree/master/examples/deep_fsa}.

\section{Stimuli for Simulation 1}
The speech sequences for Simulation 1 were synthesized using GnuSpeech using the following procedure.

We started by generating 12 text strings that were random combinations of the 12 consonant-vowel (CV) pairs . An additional random CV was added to the beginning and end of each string. Each CV other than the last was followed by a comma and a space to produce consistent CV sounds. The two extra CVs were removed after synthesis. Here is an example:

so, pi, tu, la, do, ro, ku, bi, pa, ti, go, bu, da, ru

\bigskip
\noindent Each of the strings was passed to GnuSpeech. The settings were standard with these exceptions:
\begin{itemize}[nosep]
\item voice\_name = female
\item micro\_intonation = 0
\item macro\_intonation = 0
\item intonation\_drift = 0
\item random\_intonation = 0
\end{itemize}

\bigskip
\noindent The MainDictionary file had the following changes or additions:
\begin{itemize}[nosep]
\item bi 'b\_ah\_i \% i
\item bu 'b\_uu \% ca
\item da 'd\_ar \% a
\item do 'd\_uh\_uu \% baic
\item ku 'k\_uu \% ca
\item la 'l\_ar \% a
\item pi 'p\_ah\_i \% i
\item ro 'r\_uh\_uu \% baic
\item ti 't\_ah\_i \% i
\item tu 't\_uu \% ca
\end{itemize}

\bigskip
\noindent The resulting sound files were processed as follows:
\begin{enumerate}[nosep]
\item Each file was imported into Audacity (v2.4.2.0)
\item "Sound Finder" (Analysis Menu) was used to locate the sound between silences (-db 25.0, "minimum duration" 0.10 s, "label start" and "label end" both 0.0 s and "add label" at end set to 0 (no). We manually adjusted any "label" boundaries that seemed too inclusive or too exclusive.
\item Individual sounds were exported ("Export Multiple")
\item The xxx-01.wav and xxx-14.wav files were discarded (i.e. the extra CVs at the start and end)
\item Random instances of each CV were concatenated to form multiple instances of trisyllable words conforming to the artificial language (i.e. the four words, daropi, golatu, pabiku and tibudo)
\item Randomly chosen instances of the trisyllable words were concatenated to form the 24 possible combinations of the four words.
\item After concatenation each of the 24 wav files was opened in Audacity and the start and end of each CV was marked. This was done manually by the first author while looking at the spectrogram and listening to the speech file. The timing labels for each wav file were exported separately.
\end{enumerate}

\section{Stimuli for Simulation 2}
For the second simulation we began with a wav file of human speech (see \citet{GrafEstesLew-Williams15}). The file was about 3 minutes duration in total. The file was opened in Audacity and CV starts and ends were labeled as was done for the synthesized sounds for Simulation 1. The timing information was exported separately from the wav file.

\section{ Model Layer Sizes and Structure}

Figure \ref{fig.netview} in the main text shows the general configuration of the model. Table~\ref{tab.layer_sizes} shows the specific sizes of each of the layers, and where they receive inputs from. 

\begin{table}[h]
  \centering
\begin{tabular}{llrrlll}
\hline
     &      & \multicolumn{2}{c}{{\bf Units}} & \multicolumn{2}{c}{{\bf Pools}} & \\
{\bf Area} & {\bf Name} & {\bf X} & {\bf Y} & {\bf X} & {\bf Y} & {\bf Receiving Projections} \\
\hline
A1 & A1s & 7 & 2 & 6 & 12 &  \\
R & Rs & 7 & 2 & 6 & 12 &  \\
CB & CBs & 5 & 5 & 4 & 5 & A1s CBs RBs CBth CPBs\\
   & CBct & 5 & 5 & 4 & 5 & CBs CBct CBth CPBs CPBct \\
   & CBth & 7 & 2 & 4 & 5 & {\bf A1} CBct CPBct\\
RB & RBs & 5 & 5 & 4 & 5 & A1s Rs RBs CBs RBth RPBs\\
   & RBct & 5 & 5 & 4 & 5 & RBs RBct RBth RPBs RPBct \\
   & RBth & 7 & 2 & 4 & 5 &  {\bf R} RBct RPBct\\
CPB & CPBs & 5 & 5 & 3 & 5 & CBs CPBs RPBs CPBth STSs\\
   & CPBct & 5 & 5 & 3 & 5 & CPBs CPBct CPBth STSct \\
   & CPBth & 7 & 2 & 3 & 5 &  {\bf A1} CPBct STSct\\
RB & RBs & 5 & 5 & 3 & 5 & RBs RPBs CPBs RPBth STSs\\
   & RBct & 5 & 5 & 3 & 5 & RPBs RPBct RPBth STSct \\
   & RBth & 7 & 2 & 3 & 5 &  {\bf R} RPBct STSct\\
STS & STSs & 5 & 5 & 3 & 5 & CPBs RPBs STSs STSth\\
   & STSct & 5 & 5 & 3 & 5 & STSs STSct STSth \\
   & STSth & 7 & 4 & 3 & 5 &  {\bf A1 \bf R} CPBct RPBct \\
\hline
\end{tabular}
\caption{\footnotesize Layer sizes, showing numbers of units in one pool and the number of Pools of such units, along X,Y axes.  Each area, other than A1 and R, has three associated layers: {\em s} = superficial layer, {\em ct} = corticothalamic layer (context updated by 51B neurons in same area, shown in bold), {\em th} = thalamic layer (driven by 5IB neurons from associated area, shown in bold).}
\label{tab.layer_sizes}
\end{table}

\section{Additional Results}

Figures showing results for each simulation when the model is \emph{not} first trained on the TIMIT sentences.

\begin{figure}[h]
  \centering\includegraphics[width=6in]{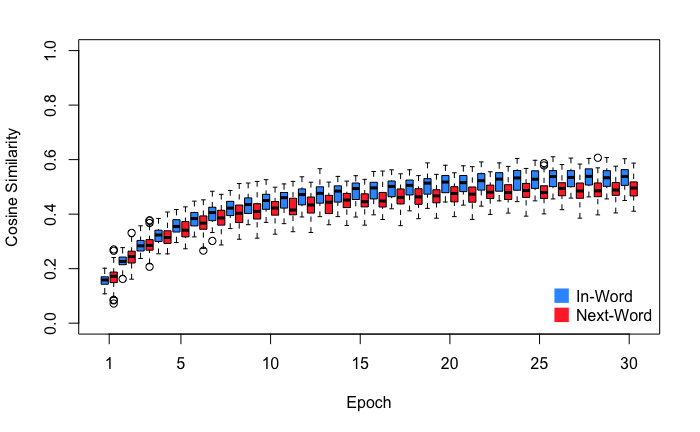}
  \caption{\footnotesize A box plot of testing results for Simulation 1 on a network that was not pretrained on the TIMIT sentences. The model learns to predict to a similar degree as when pretrained but requires more epochs of training. Testing was done after each epoch during training. The measure is the similarity of activations at the end of the \enquote{minus} phase of each trial (prediction) and the actual neuron activations at the end of the \enquote{plus} phase. Compare with Figure \ref{fig.saffranPrePostTesting}.}
  \label{fig.saffranNoPretrainTesting}
\end{figure}

\begin{figure}[h]
  \centering\includegraphics[width=6in]{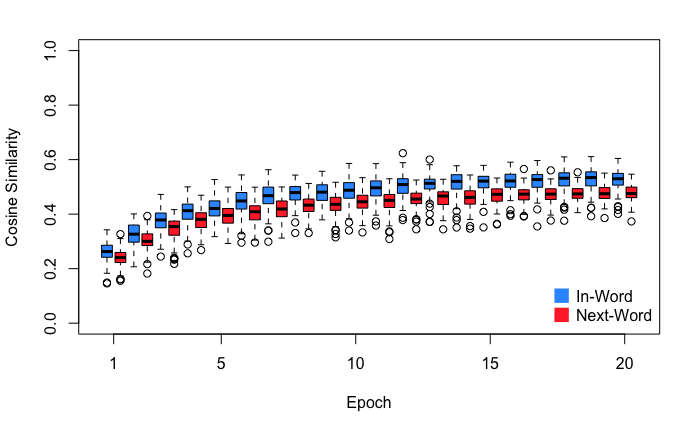}
  \caption{\footnotesize A box plot of testing results for Simulation 2 on a network that was not pretrained on the TIMIT sentences. The model learns to predict to a similar degree as when pretrained but requires more epochs of training. Testing was done after each epoch during training. The measure is the similarity of activations at the end of the \enquote{minus} phase of each trial (prediction) and the actual neuron activations at the end of the \enquote{plus} phase. Compare with Figure \ref{fig.grafestesPrePostTesting}.}
  \label{fig.grafestesNoPretrainTesting}
\end{figure}

\clearpage

\bibliography{bib/wordseg.bib}

\end{document}